# Kinematic and Dynamic Forcing Strategies for Predicting the Transport of Inertial Capsules via a combined Lattice Boltzmann – Immersed Boundary Method

A. Coclite[a], S. Ranaldo[b], M. D. de Tullio[b], P. Decuzzi[a,♣], G. Pascazio[b, ♣*]

[a] Laboratory of Nanotechnology for Precision Medicine, nPMed, Fondazione Istituto Italiano di Tecnologia, Via Morego 30-16163, Genova, Italy.

[b] Dipartimento di Meccanica, Matematica e Management, DMMM, Politecnico di Bari, Via Re David, 200-70125, Bari, Italy.

♣ PD and GP share the senior authorship.

*Corresponding author: giuseppe.pascazio@poliba.it

**ABSTRACT**

Modeling the transport of deformable capsules under different flow regimens is crucial in a variety of fields, including oil rheology, blood flow and the dispersion of pollutants. The aim of this study is twofold. Firstly, a combined Lattice Boltzmann – Immersed Boundary (LBM – IB) approach is developed for predicting the transport of inertial deformable capsules. A Moving Least Squares (MLS) scheme has been implemented to correlate the pressure, velocity and force fields of the fluid domain with the capsule dynamics. This computational strategy has been named LBM – *Dynamic IB*. Secondly, this strategy is directly compared with a more conventional approach, named LBM - *Kinematic IB*, where capsules move with the same velocity of the surrounding fluid. Multiple test cases have been considered for assessing the accuracy and efficiency of the *Dynamic* over *Kinematic IB* scheme, including the stretching of circular capsules in shear flow, the transport in a plane Poiseuille flow of circular and biconcave capsules, with and without inertia. By monitoring the capsule geometry over time, the two schemes have been documented to be in excellent agreement, especially for low Capillary numbers ($Ca \leq 10^{-2}$), in the case of non-inertial capsules. Despite a moderate increase in computational burden, the presented LBM – *Dynamic* IB scheme is the sole capable of predicting the dynamics of both non-inertial and inertial deformable capsules. The proposed approach can be efficiently employed for studying the transport of blood cells, cancer cells and nano/micro capsules within a capillary flow.

# 1. INTRODUCTION

The coupling between fluid and structure dynamics is of great relevance in different disciplines. Biophysicists are investing increasingly more efforts into modeling the flow of complex fluids, such as whole blood, to better understand the mechanisms underlying the development of diseases and their possible cure. [1-4] In a wide range of engineering problems, there is a growing demand to investigate the rheology of active fluids, oils, polymeric suspension, or colloidal mixtures moving into tortuous channels, with either fixed or variable geometries. This is specifically the case of enhanced oil recovery, trickle bed reactors, and microfluidics devices. [5-12] Regardless of the application, it is easy to recognize that immersed structures may be dragged away downstream by large distances, far from their original locations, or significantly deformed due to an incoming flux. On this premise, it is crucial to have access to computational tools capable of efficiently handle the variation in time of immersed geometries without any loss of accuracy. [13-16]

One of the most suitable computational technique to deal with the motion and deformation of particles into fluids is the Immersed Boundary (IB) method, which was originally developed by Peskin in 1972 to simulate cardiac mechanics [17]. This technique prescribes the evolution of the fluid on an Eulerian Cartesian grid, which is not conforming to the geometry of the heart, while a transferring function imposes the effect of the moving walls onto the flow field. Since then, several scientists employed this technique in virtue of its simplicity and efficiency, as an alternative to more conventional computational schemes requiring body fitted meshes. [18, 19] Indeed, the Peskin technique offers the great advantages of avoiding the mesh generation step, which is particularly expensive in the case of moving bodies where remeshing is required at each step following the immersed structure deformation.

Recently, in the IB context, two different computational schemes have been proposed to enforce the boundary conditions on the immersed structure. On one hand, no-slip conditions can be exactly imposed by advecting the Lagrangian points of the immersed surface with the underlying fluid velocity; while the effect of the body on the surrounding fluid is subsequently taken into account by means of a forcing term, which is obtained by the solid boundary model and added into the momentum equations for the fluid. In the sequel, this approach is referred to as the *Kinematic IB* [20-23]. On the other hand, the immersed body motion can be determined by the contribution of the hydrodynamic external forces and internal structural forces by solving the Newton second law. While the no-slip condition is enforced by introducing an effective forcing term, which directly acts on the surrounding fluid and accounts for the presence of the boundary. In the sequel, this approach is referred to as the *Dynamic IB* [24-27]. This approach has been already effectively used by the authors to study the transport of rigid particles with arbitrary shapes [24, 28, 29].

In this work, the authors have employed and critically compared the *Dynamic IB* and *Kinematic IB* schemes, coupled with a Lattice Boltzmann-BGK method (LB), for describing the transport of inertial deformable capsules immersed in a fluid. Precisely, the LB-BGK equation [30-32] is used for describing the fluid, while the Moving Least Squares (MLS) reconstruction by Liu and Gu [33] is used to accurately interpolate the pressure, velocity and forcing fields between the Eulerian and Lagrangian points.

Here, the authors have extended the approach to the case of an incompressible fluid interacting with deformable capsules. Both, the *Dynamic IB* and *Kinematic IB* schemes are quantitatively validated against well-known benchmark tests. First, the deformation of a circular capsule in shear flow, considering different values of the nondimensional shear resistance (G=0.0125, 0.04, and 0.125) and four values of the nondimensional bending modulus ($E_b$=0.0, 0.025, 0.05, 0.1, and 0.2). Then, the transport of circular and biconcave capsules in a plane-Poiseuille flow is studied with the capillary number, Ca, varying over three orders of magnitude ($10^{-4} \leq Ca \leq 10^{-1}$). Additionally, the transport of a single-file red blood cell-shaped capsules is considered to investigate the model behavior in the case of multi-particle transport. Finally, the effect of the inertia is studied transporting an isolated circular capsule in a linear flow and in a plane-Poiseuille flow.

## 2. METHOD

***2.1 The lattice Boltzmann method.*** The evolution of the fluid is defined in terms of a set of $N = 9$ discrete distribution functions $[f_i]$, $(i = 0, \ldots, 8)$, which obey the two-dimensional Boltzmann equation

$$f_i(\boldsymbol{x} + \boldsymbol{e_i}\Delta t, t + \Delta t) - f_i(\boldsymbol{x}, t) = -\frac{\Delta t}{\tau}\left[f_i(\boldsymbol{x}, t) - f_i^{eq}(\boldsymbol{x}, t)\right], \quad (1)$$

in which $\boldsymbol{x}$ and $t$ are the spatial and time coordinates, respectively; $[\boldsymbol{e_i}]$, ($i = 0,\ldots,8$) is the set of discrete velocities; $\Delta t$ is the time step; and $\tau$ is the relaxation time given by the unique non-null eigenvalue of the collision term in the BGK-approximation [30]. The kinematic viscosity, $\upsilon$, is strictly related to the single relaxation time $\tau$ as $\upsilon = c_s^2\left(\tau - \frac{1}{2}\right)\Delta t$ being $c_s = \frac{1}{\sqrt{3}}\frac{\Delta x}{\Delta t}$ the reticular speed of sound. The moments of the distribution functions define the fluid density, $\rho = \sum_i f_i$, velocity, $\boldsymbol{u} = \sum_i f_i \boldsymbol{e_i}/\rho$, and the pressure, $p = c_s^2 \rho = c_s^2 \sum_i f_i$. The local equilibrium density functions $[f_i^{eq}]$ ($i = 0,\ldots,8$) are expressed by the Maxwell-Boltzmann distribution,

$$f_i^{eq}(\boldsymbol{x},t) = \omega_i \rho \left[1 + \frac{1}{c_s^2}(\boldsymbol{e_i}\cdot\boldsymbol{u}) + \frac{1}{2c_s^4}(\boldsymbol{e_i}\cdot\boldsymbol{u})^2 - \frac{1}{2c_s^2}\boldsymbol{u}^2\right]. \quad (2)$$

The set of discrete velocities is given by: [34]

$$\boldsymbol{e_i} = \begin{cases} (0,0), & if\ i = 0 \\ \left(\cos\left(\frac{(i-1)\pi}{2}\right), \sin\left(\frac{(i-1)\pi}{2}\right)\right), & if\ i = 1 - 4 \\ \sqrt{2}\left(\cos\left(\frac{(2i-9)\pi}{4}\right), \sin\left(\frac{(2i-9)\pi}{4}\right)\right), & if\ i = 5 - 8 \end{cases}, \quad (3)$$

with the weights: $\omega_i = 1/9$ for i = 1−4, $\omega_i = 1/36$ for i = 5−8, and $\omega_0 = 4/9$. Here, the discretization in the velocity space for the equilibrium distribution functions is based on the quadrature of the Hermite polynomial expansion of this distribution. [24, 35] A convenient forcing term ($\mathcal{F}_i$) accounts for the presence of an arbitrary shaped body immersed into the regular Cartesian lattice, as described in the following sub-section;

external boundaries of the computational domain are treated with the known-velocity bounce back conditions by Zou and He [36].

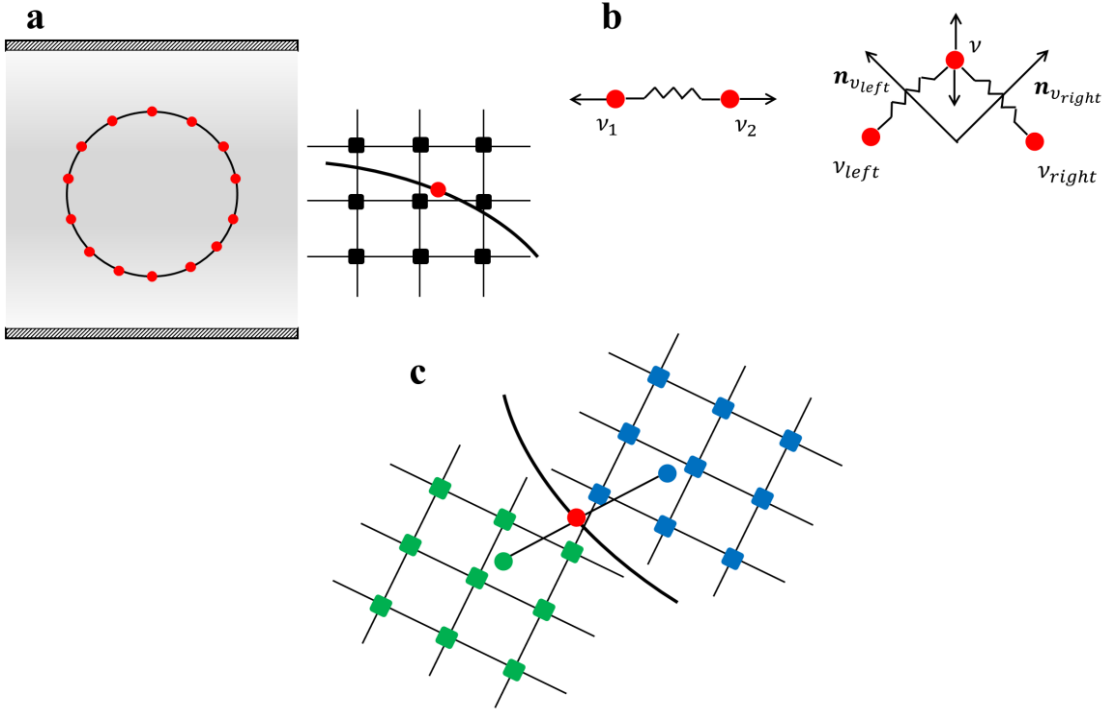

**Figure.1: Immersed boundary technique for deforming structures. a.** Schematic of the Lagrangian markers (red circles) defining the immersed structure and sketch of the nine Eulerian points (black squares) composing the support domain associated with the selected Lagrangian marker. **b.** Scheme of the elastic potentials acting on the particle's vertices. **c.** Support domains centered on the outward (green circle) and inward (blue circle) probes used to compute hydrodynamics stresses.

*2.2 Particle-based Lagrangian structures.* An effective forcing procedure accounting for the boundary presence is employed including an additional term, $[\mathcal{F}_i]$ ($i = 0,\ldots,8$), on the right-hand side of Eq.(1):

$$f_i(\boldsymbol{x} + \boldsymbol{e}_i \Delta t, t + \Delta t) - f_i(\boldsymbol{x}, t) = -\frac{\Delta t}{\tau}\left[f_i(\boldsymbol{x},t) - f_i^{eq}(\boldsymbol{x},t)\right] + \Delta t \mathcal{F}_i. \qquad (4)$$

$\boldsymbol{\mathcal{F}}_i$ is expanded in term of the reticular Mach number, $\frac{e_i}{c_s}$, resulting:

$$\boldsymbol{\mathcal{F}}_i = \left(1 - \frac{1}{2\tau}\right)\omega_i \left[\frac{\boldsymbol{e}_i - \boldsymbol{u}}{c_s^2} + \frac{\boldsymbol{e}_i \cdot \boldsymbol{u}}{c_s^4}\boldsymbol{e}_i\right] \cdot \boldsymbol{f}_{ib}, \qquad (5)$$

where $\boldsymbol{f}_{ib}$ is a convenient body force term. Due to the presence of the forcing term the macroscopic quantities are now obtained as:

$$\rho = \sum_i f_i, \qquad (6)$$

$$\rho \boldsymbol{u} = \sum_i f_i \boldsymbol{e}_i + \frac{\Delta t}{2}\boldsymbol{\mathcal{F}}_i. \qquad (7)$$

Notably, in such a framework one can recover the forced Naviér–Stokes equations with second order accuracy [25, 31, 37-39].

The forcing term is evaluated according to an Immersed-Boundary technique [17, 20, 40]. Referring to **Figure.1a**, the surface of the immersed particle is discretized via a collection of Lagrangian markers (red circles), which are superimposed onto the withstanding Eulerian fluid lattice (black squares). The forcing term in the governing equation, Eq.(4), is evaluated according to the moving–least-squares (MLS) reconstruction [26] as proposed by Coclite et al. [24]. For each Lagrangian marker (with index l), a 2D support domain is identified by nine Eulerian points arranged in a regular square, with side equal to $r_w = 2.6\,\Delta x$, and centered on the Eulerian point closest to the Lagrangian marker itself (**Figure.1a**). Given the solution at time $n$, the velocity of the Lagrangian marker is evaluated as,

$$\boldsymbol{U}(\boldsymbol{x}) = \sum_{k=1}^{9} \phi_k^l(\boldsymbol{x})\,\boldsymbol{u}_k, \qquad (8)$$

where $\boldsymbol{u}_k$ indicates the velocity at the k-th Eulerian point of the support domain and $\phi$ is the transfer operator obtained minimizing with respect to $\mathbf{a}(\mathbf{x})$ the following weighted $L_2$-norm:

$$J = \sum_{k=1}^{9} W(\mathbf{x} - \mathbf{x}^k) [\mathbf{p}^T(\mathbf{x}^k) \mathbf{a}(\mathbf{x}) - \mathbf{u}_k]^2. \qquad (9)$$

In the above equation, $\mathbf{p}^T$ is a linear basis function vector, $\mathbf{a}(\mathbf{x})$ a vector of coefficients such that $\sum_{k=1}^{9} \phi_k^l(\mathbf{x}) \mathbf{u}_k = \mathbf{p}^T(\mathbf{x}^k) \mathbf{a}(\mathbf{x})$, $\mathbf{x}$ the marker position, and an exponential weight function has been used, $W(\mathbf{x} - \mathbf{x}^k) = e^{-(\frac{r_k}{\alpha})^2}$ with $r_k = \frac{|\mathbf{x} - \mathbf{x}^k|}{r_w}$ and $\alpha=0.3$.

In this work, the coupled dynamics of deformable particles is considered for simulating the behavior of bodies, such as vesicles and membranes, effectively modeled by a closed structural surface separating the internal fluid from the external one. Deformable particles models are commonly based on continuum approaches using strain energy functions to compute the membrane response [21, 41, 42]. However, a particle-based mesoscopic model has emerged that is able to provide consistent predictions while maintaining a simple mathematical form [43-45]. The structure interacts with the surrounding (internal and external) fluids and the resulting fluid-structure interaction model is crucial for predicting the correct coupled dynamics. In fact, the particle moves according to the external forces exerted by the fluid and to the internal forces, whereas the particle position defines the fluid dynamics by means of the boundary condition to be imposed on the particle surface. Here, the immersed structure consists in a network of $nv$ vertices linked with $nl$ linear elements, whose centroids are identified with the *Lagrangian markers*.

*2.2.1 Elastic Membrane deformation.* The membrane deformations are subject to elastic strain response, bending resistance, and total enclosed area conservation.

The stretching elastic potential acting on the two vertices sharing the $l$-th element is:

$$V_l^s = \frac{1}{2} k_s (l_l - l_{l,0})^2, \qquad (10)$$

being $k_s$ the characteristics elastic constant, $l_l$ the current length of the $l$-th element, and $l_{l,0}$ the length of the $l$-th element in the stress-free configuration. Taking the derivative of the potential energy with respect to the displacements, the nodal forces corresponding to the elastic energy for nodes $v_1$ and $v_2$ (see **Figure.1b**) connected by the edge $l$ read:

$$\begin{cases} \boldsymbol{F}_{v_1}^s = -k_s (l_l - l_{l,0}) \frac{\boldsymbol{r}_{1,2}}{l} \\ \boldsymbol{F}_{v_2}^s = -k_s (l_l - l_{l,0}) \frac{\boldsymbol{r}_{2,1}}{l} \end{cases}, \qquad (11)$$

where $\boldsymbol{r}_{i,j} = \boldsymbol{r}_i - \boldsymbol{r}_j$ with $\boldsymbol{r}_i$ is the position vector of node $i$. The membrane elastic constant is computed through the capillary number, $\mathrm{Ca} = \frac{v_{ref}\, \rho_{ref}\, u_{max}}{k_s}$, or equivalently through the dimensionless shear rate, $\mathrm{G} = \frac{v_{ref}\, \rho_{ref}\, \gamma\, d_{ref}}{k_s}$; where $v_{ref}$, $\rho_{ref}$, $u_{max}$, and $d_{ref}$ indicate the reference kinematic viscosity, density, velocity and length, respectively, while $\gamma$ represents the shear rate.

The bending resistance related to the $v$-th vertex connecting two adjacent elements is:

$$V_v^b = \frac{1}{2} k_b (\vartheta_v - \vartheta_{v,0})^2, \qquad (12)$$

In the above equation $k_b$ is the characteristics bending constant, whereas $\vartheta_v$ and $\vartheta_{v,0}$ indicate the angle between the two edges connected by the $v$-th vertex corresponding to the current and stress-free configurations, respectively; therefore, the nodal forces for $v_{left}$, $v$, and $v_{right}$ are:

$$\begin{cases} \boldsymbol{F}^b_{v_{left}} = k_b \sin(\vartheta_v - \vartheta_{v,0}) \dfrac{\boldsymbol{n}_{v_{left}}}{l_{left}} \\ \boldsymbol{F}^b_v = -k_b \sin(\vartheta_v - \vartheta_{v,0}) \dfrac{\boldsymbol{n}_{v_{left}} + \boldsymbol{n}_{v_{right}}}{2(l_{left}+l_{right})} \\ \boldsymbol{F}^b_{v_{right}} = k_b \sin(\vartheta_v - \vartheta_{v,0}) \dfrac{\boldsymbol{n}_{v_{right}}}{l_{right}} \end{cases}, \quad (13)$$

where $l_{right}$, $l_{left}$, $\boldsymbol{n}_{v_{left}}$, and $\boldsymbol{n}_{v_{right}}$ are the lengths and the outward normal unity vectors related to the two adjacent edges, respectively, see **Figure.1b**. In this context the relation between the strain response constant, $k_s$, and $k_b$ is expressed through their ratio and the radius of the membrane, $r$: $E_b = \dfrac{k_b}{k_s r^2}$. [46, 47]

To constraint the fluid area enclosed into the membrane to small variations with respect to its stress-free value, an effective pressure force term is considered. The penalty force is then expressed in term of the reference pressure, $p_{ref} = \rho_{ref} c_s^2$, and directed along the normal inward unity vector of the $l$-th element, $\boldsymbol{n}_l^-$, being:

$$\boldsymbol{F}^a_l = -k_a \left(1 - \dfrac{A}{A_0}\right) p_{ref}\, \boldsymbol{n}_l^-\, l_l, \quad (14)$$

with $l_l$ the length of the selected element, $k_a$ the incompressibility coefficient, $A$ the current enclosed area, $A_0$ the enclosed area in the stress-free configuration. [48] The enclosed area is computed using the Green's theorem along the curve, $A = \int x_l dy_l - y_l dx_l$. Within this formulation $k_a = 1$ returns a perfectly incompressible membrane. Note that $\boldsymbol{F}^a_l$ is evenly distributed to the two nodes connecting the $l^{th}$-element ($v_{left}$ and $v_{right}$) as, $\boldsymbol{F}^a_l = 0.5\, \boldsymbol{F}^a_{v_{left}} + 0.5\, \boldsymbol{F}^a_{v_{right}}$.

*2.2.2 Hydrodynamics stresses.* Pressure and viscous stresses exerted over a linear element are computed as:

$$F_l^p(t) = -(p_l^+ - p_l^-)\, \boldsymbol{n}_l^+\, l_l, \qquad (15)$$

$$F_l^\tau(t) = (\bar{\tau}_l^+ - \bar{\tau}_l^-) \cdot \boldsymbol{n}_l^+\, l_l, \qquad (16)$$

where $\bar{\tau}_l^+, \bar{\tau}_l^-$ and $p_l^+, p_l^-$ are the viscous stress tensor and the pressure evaluated on the element centroid from the external (+) and internal (-) fluids, respectively; $\boldsymbol{n}_l^+$ is the outward normal unit vector, see **Figure.1c**. The pressure and velocity derivatives in Eq.s (15) and (16) are evaluated considering a probe in the normal direction of each element, being the probe length $1.2\,\Delta x$, and using the cited moving least squares formulation [24]. In this framework, the velocity derivatives evaluated at the probe are considered equal to the ones on the linear element centroid as previously done by the authors [24, 26]. In this case, all force contributes are computed with respect to elements centroids and then transferred to the element vertices.

## *2.3 Dynamic Eulerian-Lagrangian coupling.*

The dynamic immersed boundary (IB) procedure proposed and validated by Coclite et al. [24], based on the tools described in the previous section 2.2, is employed to map the forcing terms from the Lagrangian markers to the Eulerian lattice so as to impose the velocity boundary condition on the structure; in this procedure the body force term in Eq.(5), $\boldsymbol{f}_{ib}$, is evaluated through the formulation by Favier et al. [49]. For completeness, the motion of the immersed body is briefly described at first for the case of a rigid body [24]; then the procedure is provided for the case of a deformable particle.

***2.3.1 Rigid structures motion.*** Rigid motion is readily obtained integrating the force and moment contributions over the particle surface, obtaining the total force, $\boldsymbol{F}^{tot}(t)$, and moment, $M^{tot}(t)$, acting on the particle from the surrounding fluid:

$$F^{tot}(t) = \sum_{l=1}^{nl}(\bar{\tau}_l^+ \cdot n_l^+ - p_l^+ n_l^+) l_l, \qquad (17)$$

$$M^{tot}(t) = \sum_{l=1}^{nl} r_l(\bar{\tau}_l^+ \cdot n_l^+ - p_l^+ n_l^+) l_l. \qquad (18)$$

Both the linear and angular accelerations are evaluated in time as, $\dot{u}(t) = \frac{F^{tot}(t)}{m}$, and $\dot{\omega}(t) = \frac{M^{tot}(t)}{I}$, where $m$ is the particle mass, and $I$ the moment of inertia. Lastly, $u(t)$ and $\omega(t)$ are computed as:

$$u(t) = \frac{2}{3}\left(2u(t-\Delta t) - \frac{1}{2}u(t-2\Delta t) + \dot{u}(t)\Delta t\right) + O(\Delta t^2), \qquad (19)$$

$$\omega(t) = \frac{2}{3}\left(2\omega(t-\Delta t) - \frac{1}{2}\omega(t-2\Delta t) + \dot{\omega}(t)\Delta t\right) + O(\Delta t^2), \qquad (20)$$

with $\Delta x = \Delta t = 1$.

***2.3.2 Deformable structures motion.*** The total force $F_v^{tot}(t)$ acting on the $v$-th vertex of the immersed surface is evaluated in time accounting for both internal, Eq.s (11), (13), and (14), and external stresses, Eq.s (15) and (16); then, the position of the vertices is updated at each Newtonian dynamic time step considering the membrane mass uniformly distributed over the $nv$ vertices [50, 51]:

$$m_v \dot{u}_v = F_v^{tot}(t) = F_v^{int}(t) + F_v^{ext}(t), \qquad (21)$$

where,

$$\begin{cases} F_v^{int}(t) = F_v^s(t) + F_v^b(t) + F_v^a(t) \\ F_v^{ext}(t) = F_v^p(t) + F_v^\tau(t) \end{cases}. \qquad (22)$$

The Verlet algorithm is employed to integrate the Newton equation of motion. Precisely, a first tentative velocity is considered into the integration process, $\dot{x}_{v,0}(t)$, obtained interpolating the fluid velocity from the surrounding lattice nodes:

$$x_v(t + \Delta t) = x_v(t) + \dot{x}_{v,0}(t)\Delta t + \frac{1}{2}\frac{F_v^{tot}(t)}{m_v}\Delta t^2 + O(\Delta t^3). \qquad (23)$$

Then, the velocity at time level $t + \Delta t$ is computed as:

$$u_v(t + \Delta t) = \frac{\frac{3}{2}x_v(t+\Delta t) - 2x_v(t) + \frac{1}{2}x_v(t-\Delta t)}{\Delta t} + O(\Delta t^2). \qquad (24)$$

Therefore, at each time step the fluid-structure dynamics is evaluated using a weak coupled approach consisting of the following sub-steps:

1. The desired velocity $\mathbf{U}_{b,l}(\mathbf{x})$ at the Lagrangian marker is imposed evaluating the forcing term $\mathbf{F}_l(\mathbf{x}) = \frac{\mathbf{U}_{b,l}(\mathbf{x}) - \mathbf{U}(\mathbf{x})}{\Delta t}$; where $\mathbf{U}(\mathbf{x})$ is given by Eq.(8).

2. For each vertex, $v$, nine Eulerian points are considered, namely the Eulerian points falling into the two-dimensional support domain, defined as a square with side equal 2.6 $\Delta$x, see **Figure.1a**;

3. The Eulerian forcing $\mathbf{f}_{ib}$ is computed collecting the terms coming from the k-th Lagrangian marker, $\mathbf{f}_{ib}^k = \sum_l c_l \, \Phi_k^l \, \mathbf{F}_l$, contributing to the Eulerian point; the scale coefficient $c_l$ is obtained by imposing the conservation of the total force acting on the fluid [26];

4. $\mathcal{F}_i$ is computed using Eq.(5);

5. The distribution functions, $f_i(\mathbf{x}, t)$, are updated using Eq. (4);

6. Macroscopic quantities, $\rho(\mathbf{x}, t)$ and $\mathbf{u}(\mathbf{x}, t)$, are updated using Eq.s (6) and (7);

7. The total force on vertex $v$, $\mathbf{F}_v^{tot}(t)$, is computed using Eq.s (11), (13), and (14), and Eq.s (15) and (16) as the sum of internal and external contributions:

$$F_v^{tot}(t) = F_v^s(t) + F_v^b(t) + F_v^a(t) + F_v^p(t) + F_v^\tau(t);$$

8. The Verlet algorithm is employed to integrate the Newton equation of motion, Eq.s (23) and (24), thus updating the particle position.

This scheme will be referred to as *Dynamic IB* in the following.

### *2.4 Kinematic Eulerian-Lagrangian coupling.*

The IB method, firstly proposed around 1970 mainly to efficiently describe the flow patterns into human heart [17, 20], has been recently adopted by a large number of scientists due to its capabilities in terms of computational efficiency and accuracy when considering membranes as dense as the surrounding fluid. [23, 52-54] Within this constraint, the structure having the same density of the fluid, one can enforce readily the no-slip condition on the immersed boundary only transferring information about the velocity field between the Lagrangian and the Eulerian grid-points. In particular, the connector between the two meshes is the convenient force term, $f_{ib}$, in Eq.(5).

The boundary conditions enforcement and the fluid-structure interaction strategy read:

i. The total force, $F_v^{tot}(t)$, is computed using only internal contributions, Eq.s (11), (13), and (14) as: $F_v^{tot}(t) = F_v^s(t) + F_v^b(t) + F_v^a(t)$.

ii. For each vertex, $v$, nine Eulerian points are considered, namely the Eulerian points falling into the two-dimensional support domain, defined as a square with side equal 2.6 $\Delta$x, see **Figure.1a**.

iii. $f_{ib}$ is obtained spreading $F_v^{tot}(t)$ into the support domain through a transferring equation, $\mathbf{f}_{ib}^k = \sum_v c_v \Phi_k^v F_v^{tot}$, where a moving least squares weighting rule is implemented [24, 55].

iv. $\mathcal{F}_i$ is computed using Eq.(5) and $f_{ib}$ of the previous step.

v. The distribution functions, $f_i(x, t)$, are updated using Eq.(4).

vi. Macroscopic quantities, $\rho(x, t)$ and $u(x, t)$ are updated using Eq.s (6) and (7).

vii. The velocity of each Lagrangian point, $\dot{x}(t)$, is computed interpolating the velocity of the surrounding fluid from the associated Eulerian points stencil.

viii. The position of each Lagrangian point is finally evaluated and updated as:

$$x(t) = \frac{2}{3}\left(2x(t - \Delta t) - \frac{1}{2}x(t - 2\Delta t) + \dot{x}(t)\Delta t\right). \quad (25)$$

This scheme will be referred to as *Kinematic IB* in the following.

**3. RESULTS AND DISCUSSION**

*3.1 Stretching of a circular capsule in shear flow.* First, the two proposed IB schemes are validated against the data published for the dynamics of a deformable capsule in a shear flow by Sui et al. [56]. Referring to **Figure.2a**, a circular capsule of diameter d is placed at the center of a square box with the side length H = 10 d. The capsule diameter is discretized by 40 $\Delta x$ and the average length of the linear elements composing the membrane boundary is equal to $0.3 \Delta x$. The imposed shear rate is $\gamma = \frac{2u_{max}}{H}$, with $u_{max}$ the velocity magnitude of the top and bottom walls, and the resulting Reynolds number is $\text{Re} = \frac{\gamma d^2}{\upsilon}$ (= 0.05). The behavior of the capsule is regulated by the elastic constant $k_s$, which is computed through the dimensionless shear rate $G = \frac{\upsilon_{ref}\rho_{ref}\gamma d}{k_s}$, and by the bending resistance modulus, $E_b = \frac{k_b}{k_s r^2}$. Lastly, the mass of the capsule is determined considering the solid vesicles as dense as the surrounding fluid, $\frac{\rho_s}{\rho_f} = 1$. The flow field is initialized with a linear velocity profile given by $u_x(t=0) = \gamma(y - H/2) = 2u_{max}(y - H/2)/H$, $u_y(t=0) = 0$.

The accuracy of the two IB schemes is characterized by computing the time variation of the Taylor deformation parameter, $D_{xy} = \frac{a-b}{a+b}$, where $a$ and $b$ correspond to the major and minor radii of the capsule. This parameter is plotted in **Figure.2b** for $E_b = 0$, G = 0.0125, 0.04, and 0.125. The solid lines are for the *Dynamic IB* scheme, the dashed lines are for the *Kinematic IB* scheme, the black dots are for the benchmark data by Sui et al. [56]. An excellent agreement is found for the smaller shear rates (G=0.0125 and G=0.04). The relative error with respect to benchmark data $\left(\varepsilon_{Dyn/Kin} = \frac{D_{xy,Dyn/Kin} - D_{xy,Sui}}{D_{xy,Sui}}\right)$ is smaller than 1% for both schemes (**Figure.2c**). Note that the slight increase in relative error with the nondimensional shear rate is due to the constitutive model used for the capsule membrane. Indeed, it is well known that a spring model is accurate for small deformations with respect to the original configuration. [52] Then, the bending resistance of the capsule is varied ($E_b$ = 0, 0.025, 0.05, 0.1 and 0.2) having fixed G = 0.04. Results are provided in **Figure.2d**, and the corresponding relative error is reported in **Figure.2e**. Even in this case, the relative error is well confined within 1%. Given the weak coupling between the structural and fluid solution in time, the results for capsules with stiffer membranes may suffer of numerical instabilities. Indeed, the fluid may experience large displacement gradients due to the elastic response of relatively rigid capsules (see $E_b$ = 0.2 in **Figure.2d**). It is noteworthy that the *Dynamic IB*, which is well suited for $\frac{\rho_s}{\rho_f} > 1$, can accurately reproduce the benchmark results even in the case of $\frac{\rho_s}{\rho_f} = 1$. The equilibrium configurations of the capsules obtained by the *Dynamic IB* and *Kinematic IB* are drawn in **Figure.2f** for $E_b$ = 0, G = 0.0125, 0.04, and 0.125. Collectively, these data indicate that the two IB schemes return almost identical results within the tested ranges of G and $E_b$. Importantly, for the case in hand, the computational time needed by the *Dynamic IB* scheme to perform 10,000 time steps was about 7% higher than that required by the *Kinematic IB* scheme, with a computational domain of 4,000,000 points. The overall computing time, determined via "mpi_wtime", for this test case is 45900.3 sec, using 4 CPUs (Intel Xeon E5-2660 v3 2.60 GHz).

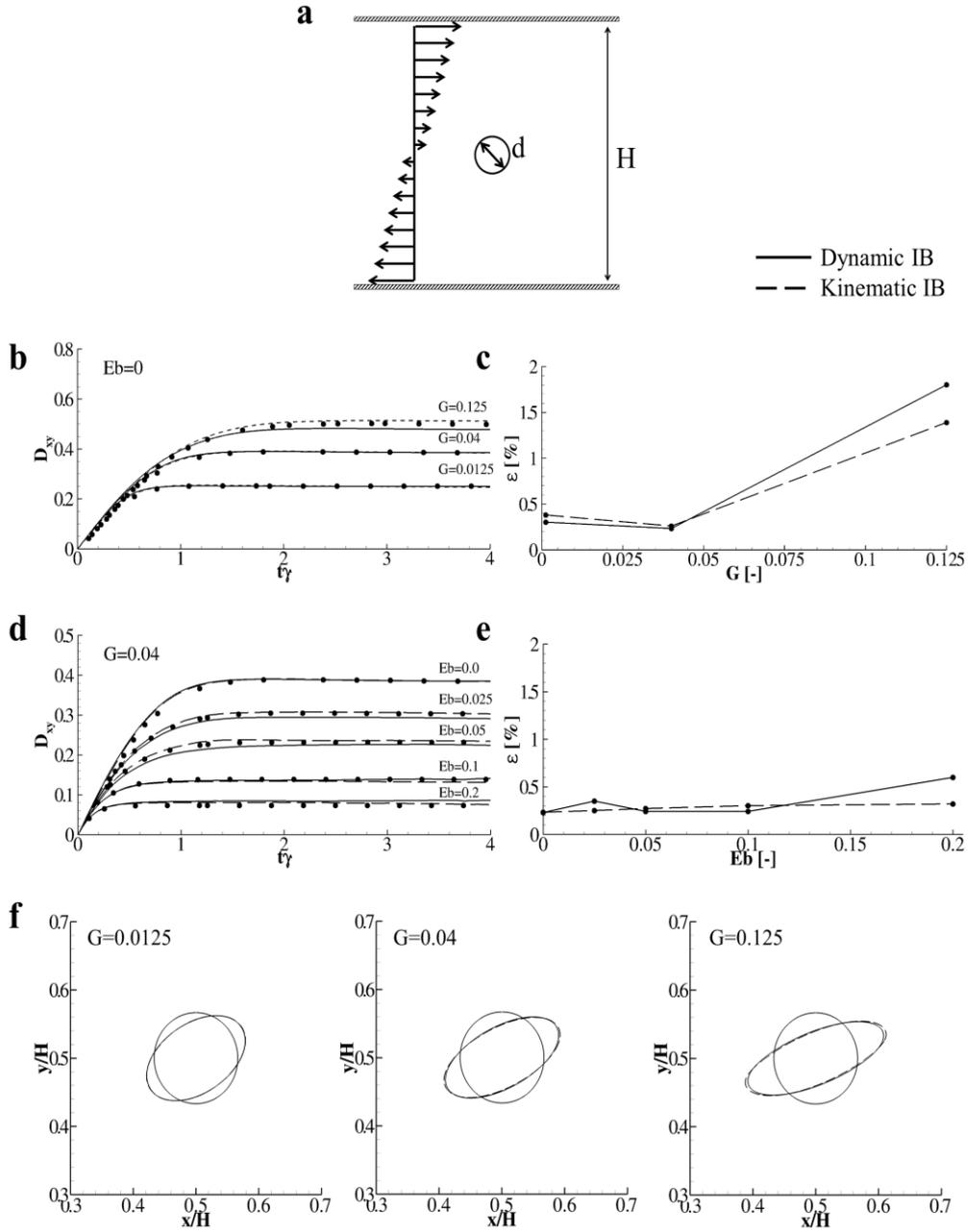

**Figure.2: Deformation of a circular capsule in shear flow. a.** Schematic representation of a circular capsule under linear shear flow. **b.** Variation of the Taylor parameter over time as function of the dimensionless shear rate G ($E_b = 0$). **c.** Relative error (%) with respect to the benchmark data [52] obtained by varying G for $E_b = 0$. **d.** Variation of the Taylor parameter over time as function of the bending stiffness $E_b$ (G = 0.04). **e.** Relative error (%) with respect to the benchmark data [52] obtained by varying $E_b$ for G = 0.04. **f.** Configurations of circular capsules for Eb = 0 and G=0.0125, 0.04, and 0.125 as compared to the initial unperturbed configuration (circle).

(Solid line: *Dynamic IB*; Dashed line: *Kinematic IB*; Symbols: benchmark data by Sui et al. [52])

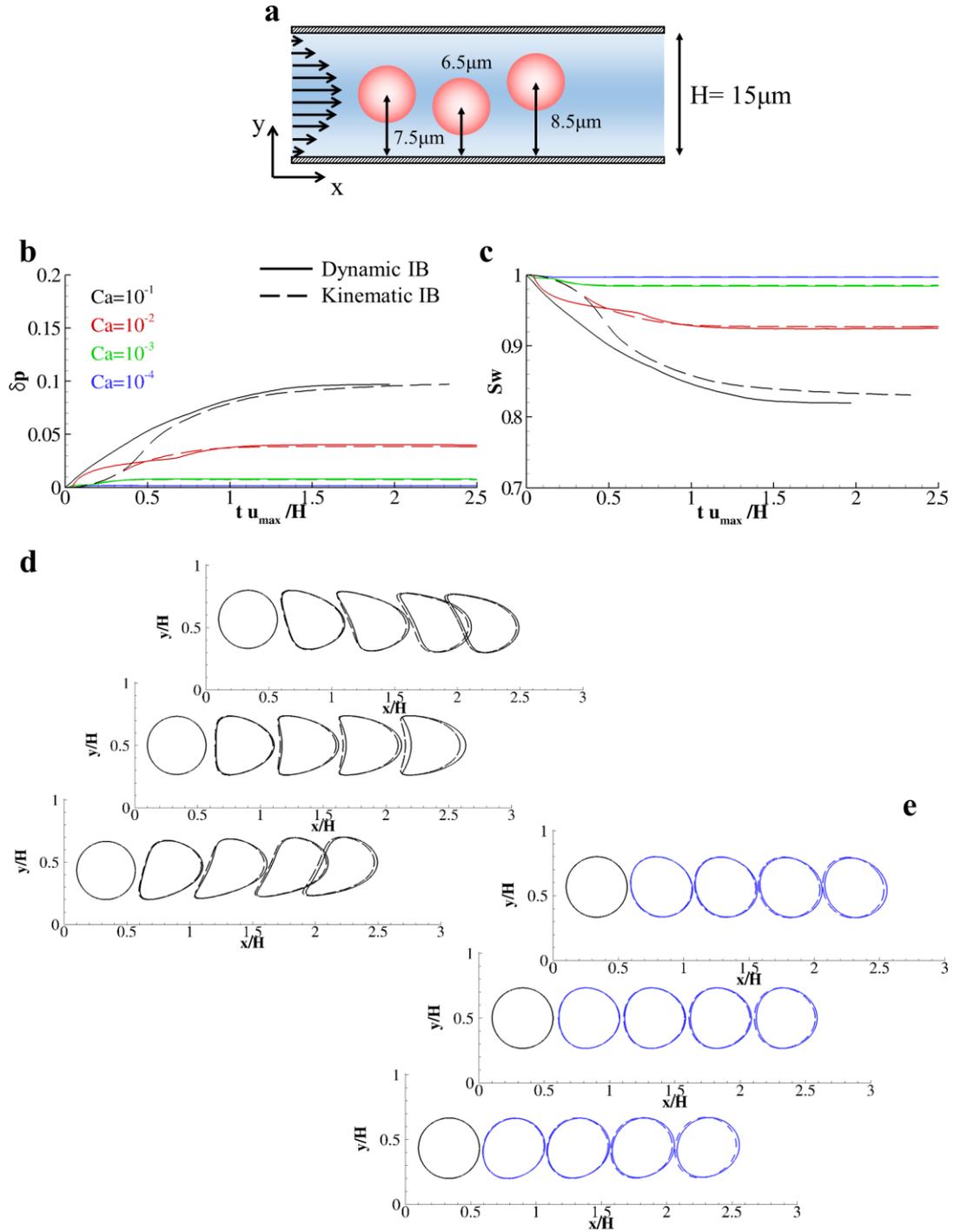

**Figure.3: Transport of a circular capsule in plane-Poiseuille flow. a.** Schematic representation of circular capsules in a rectangular channel located at different distances from the bottom wall. **b.** Variation of the capsule relative perimeter over time as function of the capillary number Ca. **c.** Variation of the capsule swelling ratio over time as function of the capillary number Ca. **(d,e).** Capsules for $Ca = 10^{-1}$ **(d)** and $Ca = 10^{-4}$ **(e)** taken at $\frac{tu_{max}}{H} = 0, 0.5, 1.0, 1.5, 2.0$.

(Solid line: *Dynamic IB*; Dashed line: *Kinematic IB*)

## 3.2 Transport of a circular capsule in a plane-Poiseuille flow.

In this second test case, the dynamics of a capsule with $\frac{\rho_s}{\rho_f} = 1$ within a plane-Poiseuille flow is considered. The capsule has a diameter of $7\ \mu m$ and is immersed in a 2D channel with a height $H = 15\ \mu m$ and length equal $3H$. A sketch of the physical problem is drawn in **Figure.3a**. Three different initial positions of the capsule within the channel are considered, whereby the center of the capsule can be at 6.5, 7.5 and 8.5 μm away from the lower wall. The capsule surface is discretized with a network of linear elements with an average length equal to $0.3\ \Delta x$. Simulations are run at $Re = 0.01$ and with four capillary numbers, namely $Ca = \frac{u_{max} v_{ref}}{k_s} = 10^{-1}, 10^{-2}, 10^{-3}$, and $10^{-4}$. In the initial condition, the flow field is at rest with $\boldsymbol{u} = 0$ and $p = p_{ref}$. The plane-Poiseuille flow is established by imposing a parabolic longitudinal velocity profile at the inlet section, $u_x(x = 0) = -u_{max}\left(\left|1 - \frac{y}{\frac{H}{2}}\right|^2 - 1\right)$, and a constant pressure at the outlet section, $p(x = 3H) = p_{ref}$. This, indeed, implies the realization of the plane-Poiseuille linear pressure drop ($\frac{\delta p}{\delta x} = -u_{max}\ 2\ \rho_{ref} v_{ref} \frac{4}{H^2}$) between the inlet and the outlet sections.

In this case, the comparison between the two IB schemes is provided in terms of the capsule perimeter variation with respect to its original configuration, $\delta p(t) = \frac{p(t) - p_0}{p_0}$ (**Figure.3b**), and of the swelling ratio, $Sw$, defined as the ratio between the capsule area $A(t)$ at time t and the area associated with a circle of perimeter $p$, used as reference ($Sw(t) = \frac{A(t)}{p^2(t)/4\pi}$) (**Figure.3c**). A direct comparison of the capsule membrane at different time points is also provided for the two IB schemes (**Figure.3d** and **Figure.3e**).

Even in this case, the two IB schemes are in good agreement within the small deformations range, $10^{-4} \leq Ca \leq 10^{-2}$. Quantitatively, the difference between the two schemes is computed by the relative error at $tu_{max}/H$=2.0, $\varepsilon = \frac{\delta p_{Dyn} - \delta p_{Kin}}{\delta p_{Kin}}$. This error depens on the considered capillary number: $\varepsilon(Ca = 10^{-4}) = 0.988 \times 10^{-4}$, $\varepsilon(Ca = 10^{-3}) = 3.61 \times 10^{-4}$, $\varepsilon(Ca = 10^{-2}) = 1.313 \times 10^{-3}$, and $\varepsilon(Ca = 10^{-1}) = 3.80 \times 10^{-3}$. Interestingly, the transient dynamics prescribed by the two IB schemes differs for Ca = $10^{-2}$ and $10^{-1}$ (red and gray lines in **Figures.3b**), whereas the capsule shapes almost perfectly overlap for $Ca \leq 10^{-2}$. It should be here emphasized that the hydrodynamic stresses associated with the fluid pressure field in the first instants of the simulation are small but certainly not null. Therefore, while in the *Dynamic IB* scheme capsules start to deform following the external hydrostatic pressure already at time zero, in the *Kinematic* IB schemes the initial zero velocity field results in a null velocity of the Lagrangian markers on the capsule surface, which consequently cannot feel the external pressure. This fundamental difference between the two schemes is responsible for the differences observed in capsule shapes within the initial time steps. It is worth noting that, in the case of large deformations, the authors needed to refine the Lagrangian mesh in order to ensure a Lagrangian spacing always higher than the Eulerian one during the transport dynamics of the capsule. This is strictly required for the *Dynamics IB* scheme in order to correctly impose the no-slip conditions, avoiding

"holes" on the body surface.. The linear springs network used for Ca = 10$^{-1}$ has an average length equal to $0.2\ \Delta x$.

These trends are also confirmed by analyzing the swelling ratio distributions in time. In particular, $Sw$ indicates how much the particles deviates from the circular shape ($Sw = 1$) reaching the final *bullet-like* configuration (**Figure.3c)**. Five capsule configurations, corresponding to different time instants, for $Ca = 10^{-1}$ and $10^{-4}$, are shown for both schemes in **Figure.3d** and **Figure.3e**, for three different initial locations within the channel. A good agreement is confirmed for $Ca = 10^{-4}$ (**Figure.3e**), whereas a slight difference is observed for $Ca = 10^{-1}$(**Figure.3d**), in agreement with the behavior already documented for $\delta p$ and $Sw$.

***3.3 Transport of a biconcave capsule in a plane-Poiseuille flow.*** In the same channel and with the same initial and boundary conditions of the previous test case, the behavior of biconcave capsules is also studied (**Figure.4a)**. The stress-free biconcave shape is obtained as the parametrization of the median section of a capsule, resembling a red blood cell [56, 57],

$$\begin{cases} x = a\alpha \sin q \\ y = a\frac{\alpha}{2}(0.207 + 2.003\ sin^2 q - 1.123\ sin^4 q) \cos q \end{cases}, \quad (26)$$

where $a$ is 0.122, $\alpha$ is equal to $3.5\ \mu m$, and the parameter $q$ varies within $[-0.5\pi, 1.5\pi]$. For the above parametrization, the swelling ratio in the stress-free configuration is equal to 0.481 [58]. The major axis, the initial perimeter, and the solid mesh discretization are fixed within this simulation.

Also in this case, the two IB schemes predict a similar behavior (**Figure.4b** and **.4c)**. The time variation of $\delta p$ obtained with the two schemes overlaps for $Ca = 10^{-4}$ and $Ca = 10^{-3}$. Interestingly, the stretching of the biconcave capsule is about half of that obtained in the case of the circular capsule. This feature can be addressed considering the unique shape of red blood cells that would maximize the area-to-volume ratio. **Figure.4d** and **.4e** provide the qualitative comparison between the capsule shapes as predicted by the two IB schemes.

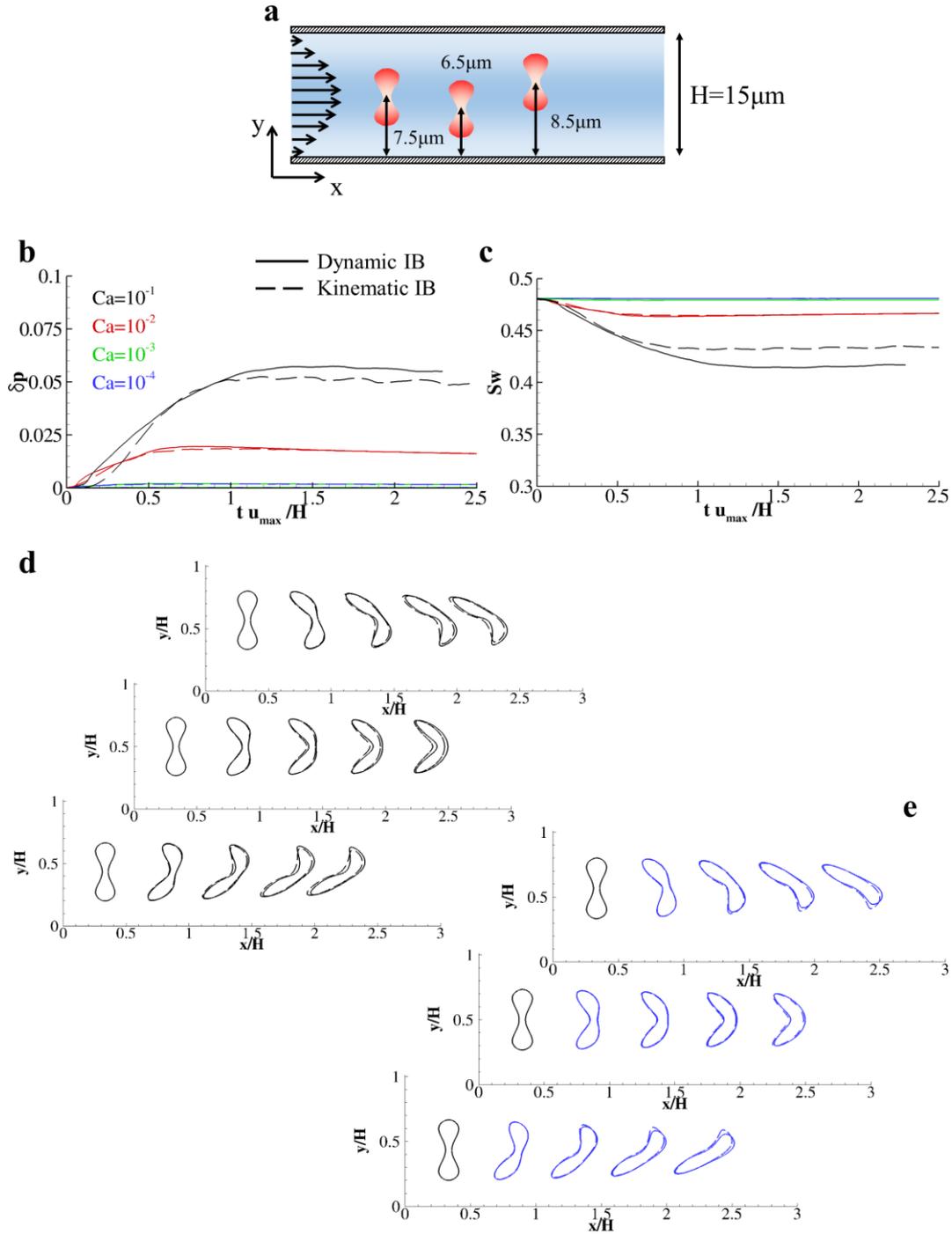

**Figure.4: Transport of a biconcave capsule in** plane-Poiseuille flow. **a.** Schematic representation of biconcave capsules in a rectangular channel located at different distances from the bottom wall. **b.** Variation of the capsule relative perimeter over time as function of the capillary number Ca. **c.** Variation of the capsule swelling ratio over time as function of the capillary number Ca. **(d,e).** Capsules for $Ca = 10^{-1}$ **(d)** and $Ca = 10^{-4}$ **(e)** taken at $\frac{tu_{max}}{H} = 0, 0.5, 1.0, 1.5, 2.0$.

(Solid line: *Dynamic IB*; Dashed line: *Kinematic IB*)

*3.4 Transport of multiple biconcave cells in plane-Poiseuille flow.* In this test case, sixteen biconcave capsules are transported in a plane-Poiseuille flow at Re=0.01 (**Figure.5a**). Collisions among the capsules are modeled through a potential field centered in each Lagrangian markers that limits to $\Delta x$ the minimum separation distance between two interacting markers. Precisely, two-body interactions are modeled through a repulsive potential centered in each vertex of the immersed particles. The repulsive force is such that the minimum allowed distance between two adjacent particles is $\Delta x$. [29] The force $F_1^{pp}$ acting on the vertex 1, at a distance $d_{1,2}$ from the vertex 2 of an adjacent particle, is directed in the inward normal direction identified by $n_1^-$ and is given by

$$F_1^{pp} = \frac{10^{-4}}{8\sqrt{2}} \sqrt{\frac{\Delta x}{d_{1,2}^5}} n_1^-. \qquad (27)$$

The elastic modulus of the capsules is $Ca = 10^{-2}$, to recover the physiological relevant conditions. The total amount of fluid enclosed by the membranes is about 30% of the total computational domain area (**Figure.5a**), thus reproducing a 30% hematocrit. Contour plots for the velocity components and pressure field at $2\,t\,u_{max}/H$ are provided for both the *Kinematic* and *Dynamic IB* schemes. A qualitative good agreement between the two schemes for the velocity fields is documented in **Figure.5b**, **.5c** and **Figure.5d**, **.5e**, for the horizontal and vertical velocity components, respectively. Some differences are instead observed for the pressure contours, as shown in **Figure.5f** and **.5g**. Precisely, the pressure fields appear quite similar when the interaction potential between cells is active (cells with other cells in their neighborhoods) and differs when the hydrodynamics alone dominates the transport (see black circles in **Figure.5f** and **.5g**).

Interestingly, three different capsule behaviors can be identified depending on the relative location of the capsule within the flow field: bulk capsule, whereby capsules are surrounded by other capsules; front capsule, referring to capsules moving at the head of the shoal; rear capsule, related to capsules moving at the back of the shoal (**Figure.6a**). As compared to bulk capsules, front and rear capsules stretch about 1.4 and 0.8% more, for both IB schemes (**Figure.6b**). Indeed, hydrodynamics stresses generated within the meniscus between two adjacent capsules locally oppose deformations.

Despite the complex dynamics depicted in this test case, the agreement between the two IB schemes continues to be confirmed. Specifically, the relative errors between the two solutions is computed using $\delta p$ and returns $\varepsilon(front) = 0.683 \times 10^{-3}$, $\varepsilon(bulk) = 1.64 \times 10^{-3}$, and $\varepsilon(rear) = 0.622 \times 10^{-3}$, depending on the capsule location. The time distribution of swelling ratio and perimeter variation show similar behaviors, drawing a picture in which, overall, bulk capsules are less stressed than the front and rear capsules (**Figure.6c**). Finally, for this test case, accounting for a domain of 540,000 Eulerian points and 16,000 Lagrangian markers distributed over 16 biconcave capsules, the computational time needed by the *Dynamic IB* scheme to perform 10,000 time steps was about 16.7% higher than that required by the *Kinematic IB* scheme. The overall computing time, determined via "mpi_wtime", for this test case is 1'000'000 sec using 27 CPUs (Intel Xeon E5-2660 v3 2.60 GHz).

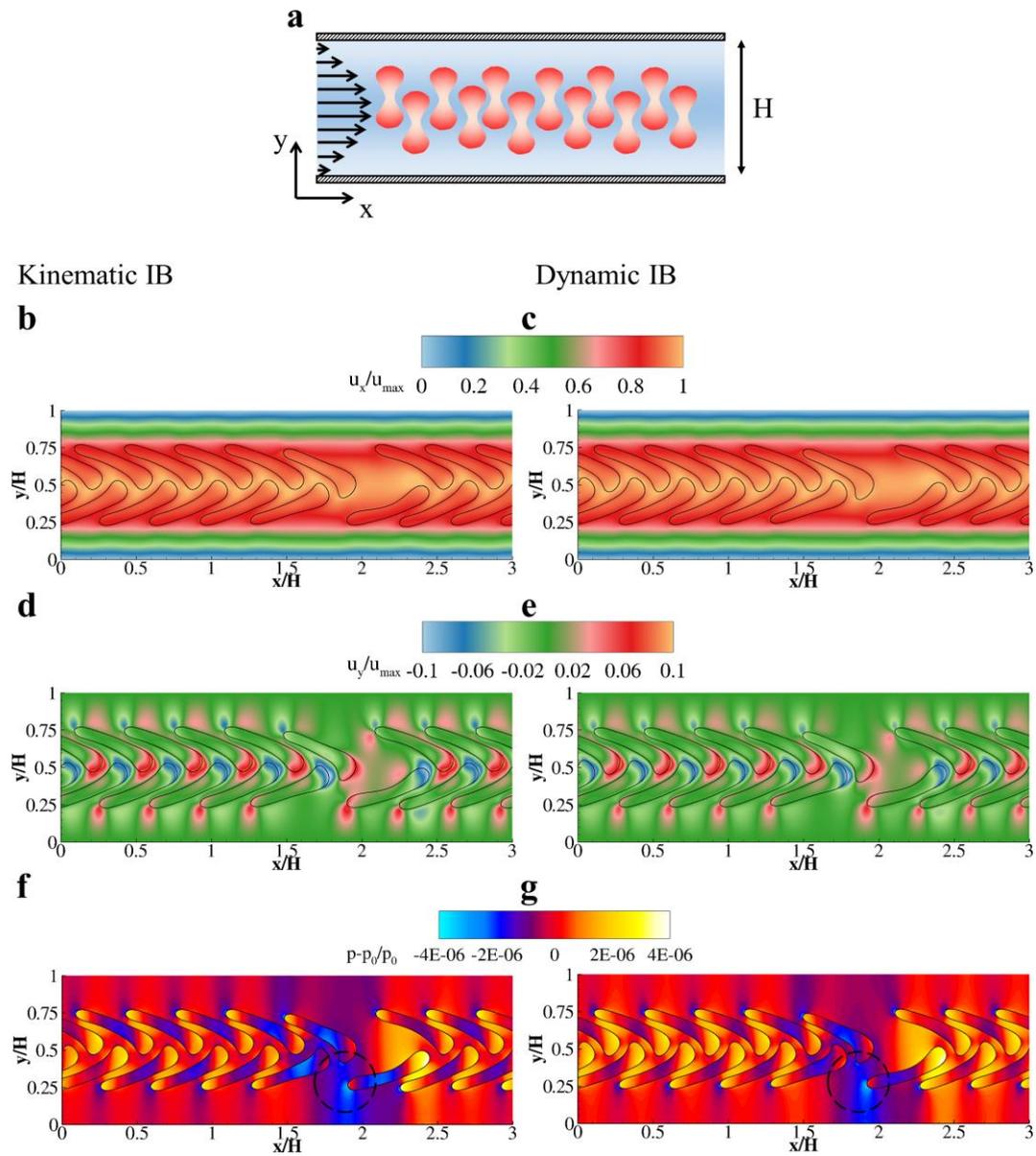

**Figure 5:** Transport of multiple biconcave capsules in a plane-Poiseuille flow. **a.** Schematic representation of biconcave capsules in the rectangular channel. **(b,c).** Contour plot of the horizontal velocity component. **(d,e).** Contour plot of the vertical velocity component. **(f,g).** Contour plot of the relative pressure with enlighten the two zones in which the two approaches differ (black dashed circles).

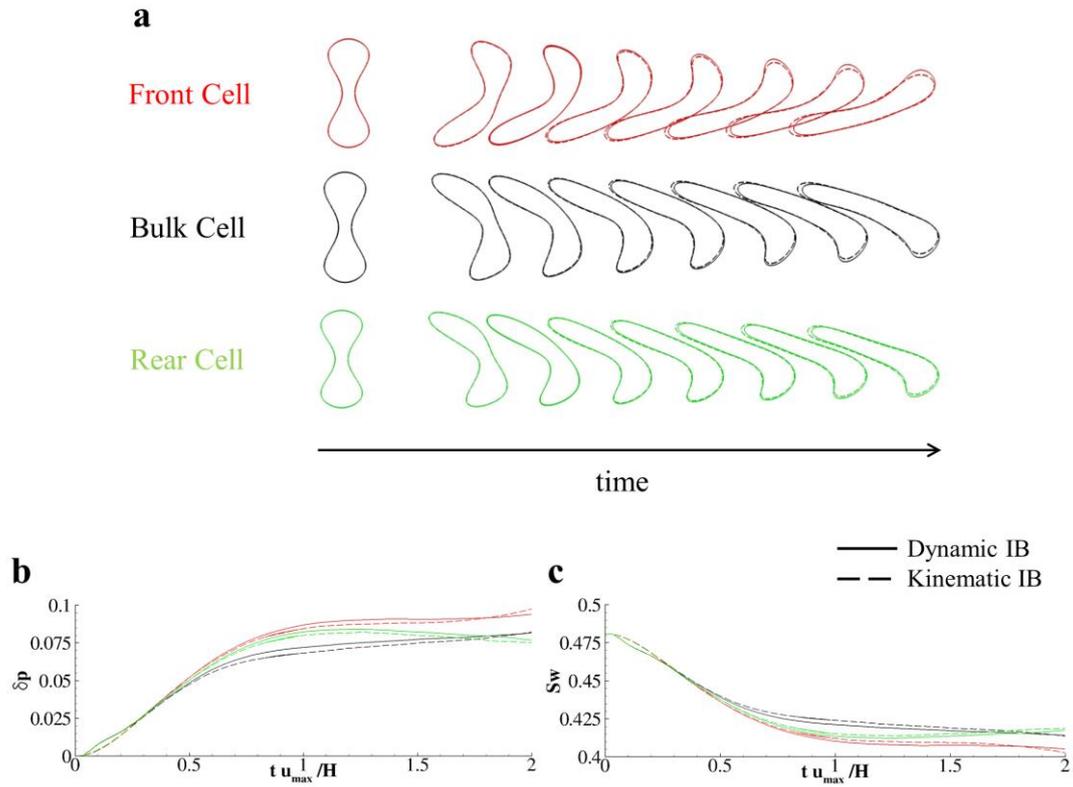

**Figure.6: Biconcave capsules regimens in Poiseuille Flows. a.** Front, Bulk, and Rear cells successive configurations for $Ca = 10^{-2}$ taken from $\frac{tu_{max}}{H} = 0$ to $\frac{tu_{max}}{H} = 2$ with non-dimensional time interval equal to 0.25. **b.** Perimeter variation distributions over time for the Front, Bulk, and Rear cells. **c.** Swelling ratio distributions over time.

(Solid lines: *Dynamic IB* scheme; Dashed lines: *Kinematic IB* scheme)

## 3.5 The role of inertia in the transport of a circular capsule in linear and parabolic flows.

The role of the inertia is assessed by computing the behavior of a circular capsule for different values of the membrane density in two different configurations: the dynamics of the capsule with Ca=10$^{-3}$, Eb=0.1 in a Couette flow (linear flow) and the detachment of a capsule from the wall in a plane-Poiseuille flow (parabolic flow). Note that this specific simulation can easily be performed using the *Dynamic IB* scheme, while the standard *Kinematic* IB scheme can solely be used for $\frac{\rho_s}{\rho_f} = 1$.

Firstly, a linear laminar flow is established in a rectangular computational domain by moving the upper wall ($y = H$) with $u_{max} = \frac{v\,Re}{H}$ (Re = 10, H = 100Δx). The diameter of the particle is d = 0.25H (**Figure.7a**) and the membrane thickness equals 0.01d. The rationale of this test case is to elucidate the role of capsule inertia as the sole independent parameter of the simulation. Indeed, an increase of the membrane mass reflects into an increase of the capsule inertia. This is expected to facilitate the lateral migration of the capsule by increasing the rotational lift component. [24]

As expected, all capsules reach the theoretical equilibrium position of 0.5H regardless of their mass as depicted in **Figure.7b**. As previously studied by the authors, the pressure field across the particle boundary is one of the components of the total lift acting on the immersed body: capsules tend to generate a symmetric pressure field across their boundary.[24] Such fields are drawn for a representative capsule with $\rho_s/\rho_f = 2$ taken at two different time instants, namely one as soon as the capsule reaches the equilibrium height (**Figure.7c**) and the other in the end of the capsule journey (**Figure.7d**). Lastly, the distributions of $\delta p$ and $Sw$ are reported in **Figure.7e** and **.7f**. Interestingly, the capsules deform almost immediately after being released and keep their deformed configuration for all the lateral migration process due to the linearity of the velocity flow field.

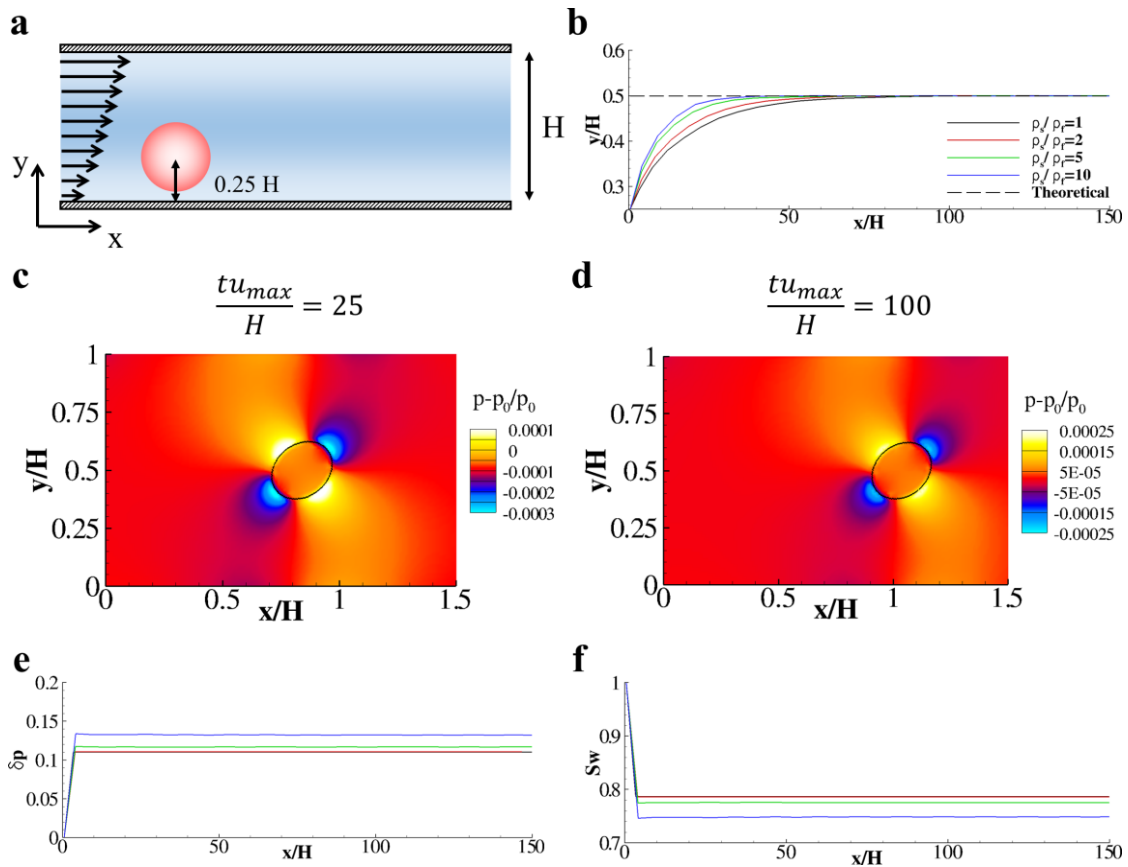

**Figure.7: The role of inertia in the transport of a circular capsule in a linear flow. a.** Schematic representation of an initially resting circular capsule in a rectangular channel with centroid at 0.25 H from the bottom wall. **b**. Trajectory of the centroid of the circular capsule for different values of the relative density $\rho_s/\rho_f$. **c.** Pressure field around the circular capsule for $\rho_s/\rho_f = 2$ taken at $\frac{tu_{max}}{H} = 25$. **d.** Pressure field around the circular capsule for $\rho_s/\rho_f = 2$ taken at $\frac{tu_{max}}{H} = 100$. **e.** Variation of the capsule relative perimeter over time as function of the relative density $\rho_s/\rho_f$. **f.** Variation of the capsule swelling ratio over time as function of the relative density $\rho_s/\rho_f$.

The role of the inertia is further demonstrated studying the detachment of a single capsule in a parabolic flow with Re=10$^{-2}$ (**Figure.8a**). In particular, as the ratio between the capsules and the surrounding fluid increases, the distributions of $\delta p$ and $Sw$ deviate more from the case of membranes having the same fluid density (**Figure.8b** and **.8c**). Indeed, a capsule would move towards different flow regions of higher velocities if it lags the flow or, in other words, if the slip velocity between the capsule and the undisturbed fluid is greater than zero. The velocity of the undisturbed flow is the velocity that the fluid would have if no particle was immersed in. Such parameter, $u_{x,slip}$, depends on the inertia moment of the immersed body, so that, in this case, on the membrane mass. Increasing the inertia causes an increase of the time is needed to equilibrate the slip velocity profile across the immersed bodies, as demonstrated in **Figure.8d**. [24]

Here, the distribution of $u_{x,slip}$ is taken at $t\frac{u_{max}}{H} = 1.5$ along the capsules boundaries by varying counterclockwise the radial coordinate θ. Notably, the larger displacement between these distributions is found in the lower region of the particles, around $\frac{3}{4}\pi$. In this zone, the particles are navigating along streamlines with lower velocity (not strong velocity gradients), while pressure stresses are large and induce lubrication lift (**Figure.8d**, **Figure.8e**). These differences are qualitatively depicted in **Figure.8f** where five membrane configurations, taken at subsequent five time instants, are presented.

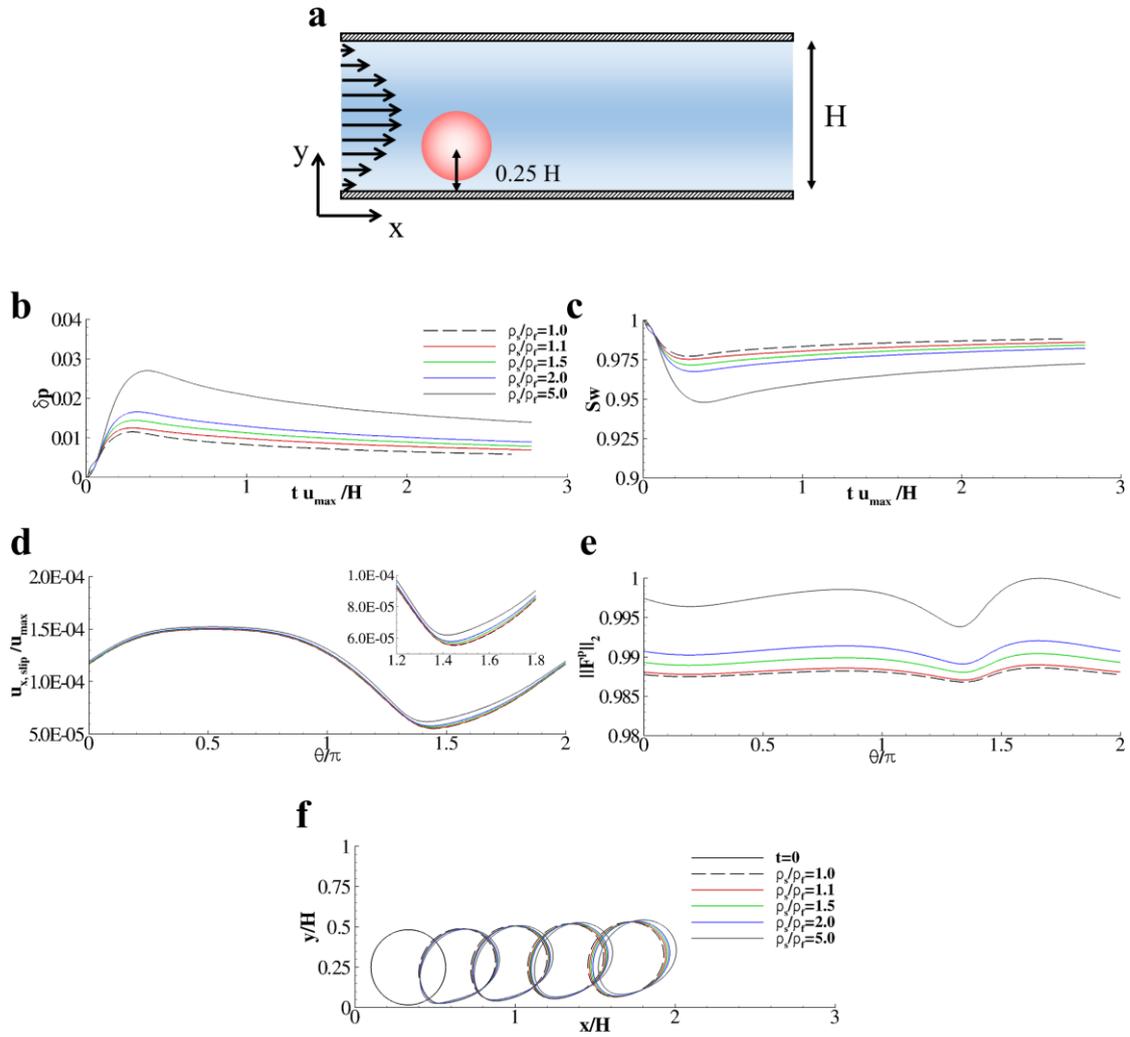

**Figure.8: The role of inertia in the transport of a circular capsule in plane-Poiseuille flow. a.** Schematic representation of circular capsules in a rectangular channel initially in contact with the bottom wall. **b.** Variation of the capsule relative perimeter over time as function of the relative density $\rho_s/\rho_f$. **c.** Variation of the capsule swelling ratio over time as function of rthe elative density $\rho_s/\rho_f$. **d.** Longitudinal slip velocity measured on the capsules perimeter at $\frac{tu_{max}}{H} = 2.0$ for different values of the relative density $\rho_s/\rho_f$. **e.** Module of the pressure force measured on the capsules perimeter at $\frac{tu_{max}}{H} = 2.0$ for different values of the relative density $\rho_s/\rho_f$. **f.** Capsules for different values of the relative density $\rho_s/\rho_f$ taken at $\frac{tu_{max}}{H} = 0, 0.5, 1.0, 1.5, 2.0$.

**CONCLUSIVE REMARKS**

A combined Lattice Boltzmann-Immersed Boundary (LB-IB) method was developed for predicting the transport of deformable capsules with arbitrary shapes within a capillary flow. A Moving Least Squares (MLS) approach was implemented to accurately interpolate the pressure, velocity and force fields between the Eulerian and Lagrangian meshes describing the fluid and the capsule dynamics, respectively. This scheme was named *Dynamic* IB. This is different from the *Kinematic* IB where the capsules move essentially with the same velocity of the surrounding fluid. The standard *Kinematic* IB is strictly valid only for neutrally buoyant capsules.

A direct comparison between the *Dynamic* and *Kinematic* IB schemes was performed considering multiple test cases, including the stretching of circular capsules in a shear flow, the transport in a plane-Poiseuille flow of circular and biconcave capsules with and without inertia. The two IB schemes were compared in terms of capsule membrane stretching and membrane shape. Across all test cases, the two schemes have provided results in excellent agreement with each other, especially for moderate deformations (low Ca numbers). Importantly, the more complex computational scheme was observed, under different conditions, to increase the computational burden as compared to the *Kinematic* IB by no more than 20%.

In summary, the proposed LB-*Dynamic* IB approach can accurately and quite efficiently predict the dynamics of deformable and rigid structures with arbitrary shapes and non-zero inertia. This approach can be applied to model the dynamics of blood cells, biological micro vesicles and particles for biomedical applications.

ACKNOWLEDGMENTS. This project was partially supported by the European Research Council, under the European Union's Seventh Framework Programme (FP7/2007-2013)/ERC grant agreement no. 616695; AIRC (Italian Association for Cancer Research) under the individual investigator grant no. 17664; European Union's Horizon 2020 research and innovation programme under the Marie Sadowska-Curie grant agreement no. 754490 "MINDED"; the Italian Institute of Technology.